\documentclass{aa}
\usepackage{epsfig}
\def \eg {{\it e.g.} }
\def \etal {{et al.} }
\def \ie {{\it i.e. } }

\def \tW {\widetilde{W}}
\def \dss {\displaystyle}
\def \enlarge {\phantom{$\Biggl($}}
\newcommand{\new}[1]{{{#1}}}

\begin{document}

\title{ Non-linear coupling of spiral waves in
disk galaxies: a numerical study}

\author{F.~Masset\inst{1,2} \and M.~Tagger\inst{1}}

\offprints{M.~Tagger}

\institute{DSM/DAPNIA/Service d'Astrophysique (URA 2052
associ\'ee au CNRS), CEA Saclay, 91191 Gif-sur-Yvette, France
\and
IRAM, Avenida Divina Pastora~7, N\'ucleo Central, 18012 Granada, Spain}
\date{Received 9 September 1996 /  Accepted 3 December 1996}
\thesaurus{322, 442--454 (1997)}
\maketitle

\begin{abstract}

We present the results of two-dimensional numerical simulations of stellar
galactic disks, aimed at studying the non-linear coupling between bar and
spiral waves and modes, in disks with realistically peaked rotation
profiles.  The power spectrum analysis of the perturbed density
in the disk, for azimuthal numbers ranging from $m=0$ to $m=4$, shows
an unambiguous signature of non-linear coupling between the bar and
spiral waves, or between spiral waves only, with a very sharp
selection
of the frequencies which optimize the coupling efficiency.  It turns out that
non-linear coupling can be quite efficient, and even more relevant
than the Swing mechanism to account for the dynamics of the galaxy
beyond the  corotation of the bar.  Non-linear coupling is also responsible
for a number of other behaviors observed in our runs, such as harmonic
or sub-harmonic excitation, and the excitation of $m=1$ spiral waves.


\end{abstract}

\section{Introduction}

The linear theory of spiral density waves and modes has been very
successful at explaining many features of the dynamics of observed or
numerically generated galactic disks.  Non-linear effects (other than
the generation of shocks in the gaseous component) are most often
discounted, because of the small relative amplitude of the waves (\ie
the ratio of the perturbed versus unperturbed density or potential):
it is usually found to be in the range~0.1--0.3, so that the quadratic
terms from which non-linearities arise are of order~0.1--0.01,
negligible compared to the linear ones (see \eg Strom \etal 1976).

If spiral waves could be described only by means
of linear processes, each wave or mode present in the disk would
behave independently; they would be subject to the direct
consequences of their linear behavior, and in particular (see Binney
\& Tremaine 1987, and references therein):
 \begin{itemize}
 	\item  they would be amplified by the Swing mechanism at
their corotation radius.

 	\item  they would conserve energy and angular momentum during their
 	propagation, except at their Lindblad resonances where they
 	exchange them with the stars by Landau effect.

	\item their radial propagation can result in transient behaviors,
	\ie waves emerging from the thermal noise or due to the tidal
	excitation of a companion, amplified once as they are reflected
	from their corotation radius, and traveling back to their
	Lindblad resonance where they are damped.

	\item they can also appear as exponentially growing normal modes:
	these result from the fact that, if a spiral wave can propagate to
	the galactic center, it is reflected towards its corotation
	radius.  There it is reflected back towards the center, at the
	same time as it is Swing-amplified.  Thus the center and
	corotation define a ``cavity'' within which the wave travels back and
	forth.  Classically, an integral phase condition then defines a
	discrete set of frequencies, such that after one propagation cycle
	the wave returns with its initial phase; since the wave is
	amplified at each cycle by the same factor (say a factor $\Gamma$
	every cycle time $\tau_c$), this defines an exponential growth
	rate $\gamma=\Gamma/\tau_{c}$.

	\item Swing amplification results from
	the property that waves inside corotation (\ie inside the cavity)
	have negative energy, associated with the fact that their
	azimuthal phase velocity (pattern speed) is lower than the
	rotation frequency of the stars.  On the other hand, as a wave is
	reflected inward, it excites beyond corotation a wave at the same
	frequency which has positive energy, because there it rotates
	faster than the stars, and which travels outward to the Outer
	Lindblad Resonance.  Conservation rules then imply that the
	negative energy of the waves inside corotation must grow at the
	rate at which positive energy is emitted beyond corotation.

  	\item  in a stellar disk, spiral waves propagate only
  	between their Lindblad resonances, where they are damped by Landau
  	effect.

\end{itemize}

The latter effect was not a strong constraint in early numerical
studies of spiral waves, which for numerical stability reasons were
done in models with very weakly peaked rotation curves at the galactic
center.  As a consequence it was easy to find normal modes with a
frequency high enough to avoid having an Inner Lindblad Resonance
(ILR) near the center, yet low enough to meet their Outer Lindblad
Resonance (OLR) only at a large radius, so that they could essentially
extend throughout the disk.  This situation changed with the
introduction of faster computers, and optimized codes which allowed to
deal with realistically peaked rotation profiles at the center.  In
that case waves can avoid crossing an ILR only if they have a high
frequency, but then their OLR occurs at a rather small radius: thus
any single wave or mode cannot efficiently carry energy and angular
momentum over a large radial span.

In that situation Sellwood (1985), who was the first to run such
realistic simulations, obtained unexpected results which were first
mentioned in his study of the spiral structure of the Galaxy; the
simulation showed two different $m=2$ (\ie two-armed) long-lived
patterns, with well-defined frequencies (typical of linear normal
modes) but which did not obey the usual rules associated with linear
theory: the inner one, a bar, had a high enough frequency to avoid an
ILR, but did not extend beyond its corotation radius located about
mid-radius of the disk, implying that it should not have been
amplified to reach its large amplitude.  The outer pattern, on the
other hand, had a lower frequency so that it did have an ILR: Landau
damping should thus have forbidden it to survive for a long time in
the disk.

Thus the simulation showed waves which followed in many respects the
behavior expected for normal modes of the linear theory, but yet
should have been forbidden by that theory; Tagger \etal (1987) and
Sygnet \etal (1988) presented an explanation based on a remark which
was later confirmed in all similar simulations: the two patterns
overlapped over a very narrow radial range, which coincided both with
the corotation of the inner one and the ILR of the outer one.  They
showed (this is in fact a generic property of non-linear wave
coupling, also observed in other fields of physics, see Tagger \&
Pellat 1982) that this
coincidence of resonances would make the non-linear interaction
between the two patterns much more efficient than the crude
order-of-magnitude estimate mentioned previously.  A detailed kinetic
description of this interaction allowed them to find that the two
patterns should exchange energy and angular momentum between them and
with their beat waves, namely an $m=4$ and an $m=0$ waves (where $m$ is the
azimuthal wave number), with respectively the sum and the
difference of their frequencies.  They found that these beat waves
also have a Lindblad resonance at the interaction radius, and that the
coincidence of these resonances does make it realistic to have a
strong exchange of energy and angular momentum, \ie a strong
non-linear effect,  even at relatively small amplitudes.

This made it possible to consider a scenario where the inner mode is
amplified by the linear Swing mechanism, and stabilized non-linearly
at a finite amplitude by transferring energy and momentum both to the
outer mode and to the coupled ones, \ie their beat waves (although the
non-linear interaction could also result in non time-steady, or even
chaotic, behaviors).  Thus the various waves involved conspire to
carry the energy and angular momentum, extracted by the first mode
from the inner parts of the disk, much farther out than it alone
could.

Unfortunately the absence in Sellwood's published runs of diagnostics
for the $m=4$ and $m=0$ modes did not allow to further check this
model.  In a series of papers Sellwood and collaborators gave a better
characterization of the behavior associated with the two $m=2$
patterns, and presented alternative explanations.  Sellwood \& Sparke
(1988), based on simulations by Sparke \& Sellwood (1987), presented
more detailed results and suggested that this behavior might be quite
common in barred spiral galaxies, where it might help solve the
long-standing conflict between the assumed locations of the
corotations of the bar and spirals.  They also pointed to an
interesting feature which might make the relation with observations
quite difficult: since the bar and the spiral rotate at different
speeds, they should at any given time be observed at random relative
azimuthal positions, \ie with the tip of the bar and the inner tip of
the spiral at the same radius, but not the same position angle (in
fact Sygnet \etal, 1988, noted that precisely this was suggested by
Sandage (1961) for SB(r) galaxies, though Buta (1987) does not confirm
this behavior).  Indeed Sellwood \& Sparke (1988) plot isocontours of
the perturbed potential, showing this property: the bar and the spiral
do form clearly detached patterns; on the other hand the isodensity
contours are much more fuzzy, and hardly ever detached: this can be
attributed to the fact that the density responds to the perturbed
potentials in a complex manner, and manages to establish material
bridges between the two patterns.  This difference might explain the
differing conclusions of Sandage and Buta, since observations deal
mainly with the density contrast (through the complex process of star
formation), and thus could not easily show the difference between the
two patterns.

Sellwood \& Kahn (1991) have considered an alternative explanation
for the behavior observed in these simulations, based on {\it
grooves} and {\it ridges} in the surface density or angular momentum
distribution.  They find numerically and analytically an instability,
which they call the ``groove mode'', and which is in fact essentially the
``negative-mass instability'' already found by Lovelace \& Hohlfeld
(1978), due to either a ridge or a groove in the density profile of
the disk.  It is composed of waves emitted on both sides of the groove
or ridge radius, and with their corotation close to that radius.

Sellwood \& Lin (1989) presented a recurrent spiral
instability cycle, based on this mechanism.  Their simulations (made
in special numerical conditions which we will discuss below) show that a
transient spiral instability can extract energy and angular momentum
from its corotation region, and transfer them to stars close to its
Lindblad resonance.  This creates a narrow feature in phase space, \ie
a groove or ridge in density space, which can give rise to a new
instability with its corotation at this radius, close to the
Lindblad resonance of the original wave.  This new instability will in
turn deposit the energy and momentum close to its Lindblad resonance,
where a third one can then develop, and so on.  In this manner
Sellwood \& Lin obtain a whole ``staircase'' of spiral patterns,
conspiring to carry the angular momentum outward, and such that each
has its corotation close to the OLR of the previous one.  They rule
out mode coupling as an explanation, by interrupting the run at a
given time, shuffling the particles azimuthally to erase any trace of
a non-axisymmetric wave, and starting over the run.  The ``scrambled''
simulation generates precisely the same wave as the original one,
proving that in these simulations coherent coupling between the waves is not
relevant.

Two main arguments make us believe that this recurrent cycle is not at
the origin of the behavior discussed in what we would call the {\it
generic} case, \ie Sellwood's (1985) work, the present one, as well as
much numerical work without the {\it ad hoc} numerical restrictions
used by Sellwood \& Lin (1989) : the first argument is that in their
cyclic mechanism the secondary wave is found to have its corotation at
the Outer Lindblad resonance of the primary; in the generic case, on
the other hand, the secondary wave is generated with its ILR at the
corotation of the primary.  Power density contours obtained in our
simulations, as will be shown below, allow to clearly discriminate
between these two possibilities, and rule out the groove
mechanism as a source for the secondary in the generic case.

Furthermore, Sellwod \& Lin use very artificial conditions to exhibit
more clearly the physics they want to discuss, in order that it does
not get blurred by all the complex, numerical or real, physics
involved in a full simulation.  In particular they make use of the
fact that their code is written in polar coordinates to artificially
eliminate any non-axisymmetric feature other than $m=4$ -- the
wavenumber of the waves considered in their work.  Thus
non-linear coupling with the $m=4 + 4 = 8$ is ruled out, making any
comparison with the generic case dubious.  Coupling with the
$m=4-4=0$, \ie the axisymmetric component,
is still possible, but then is not eliminated by the scrambling:
indeed the distribution function after scrambling has every reason not
to be an equilibrium one (\ie a function only of the constants of
motion, \ie constant along epicyclic orbits) in the region of the ILR
of the dominant wave before scrambling; thus $m=0$ oscillations at the
epicyclic frequency in the rotating frame, \ie precisely at the
frequency of the secondary wave found in the simulations, must be
generated by the scrambling, casting at least some doubts on the exact
physics of the subsequent evolution.

Thus non-linear coupling remains our preferred explanation for the
behavior of ``generic'' simulations, and in this paper we present
numerical work that substantiates this explanation.  An additional
incentive to do this is that in a more recent paper (Masset \&
Tagger, 1996b) we presented analytical work showing that non-linear
coupling is a very tempting explanation for the generation of galactic
warps: this is a long-standing problem, since warps (which are bending
waves of the disk, very similar in their physics to spiral waves) are
not linearly unstable (or extremely weakly, see Masset \& Tagger
1996a); thus, contrary to spiral waves, there is no simple way to
explain their nearly universal observation in edge-on galaxies, even
isolated ones (see Masset \& Tagger, 1996b, for a discussion of
alternative explanations that have been considered). In our mechanism
the spiral, as it reaches its OLR, can transfer to a pair of warps the
energy and angular momentum it has extracted from the inner parts of
the galactic disk.  The first warp would be the strong bending of the
gaseous disk beyond the Holmberg radius, while the second one would be
the short-wavelength corrugation observed within the Holmberg radius
in many galaxies, including ours.

In order to study this mechanism numerically, we have written a 3-D
particle-mesh code, whose results will be presented elsewhere. In a
first step, we have used a 2-D version of this code for the present
work, to give a more detailed analysis (which we consider as a
numerical confirmation) of non-linear coupling between spiral waves
and modes.
\new{Let us note that preliminary 3-D runs already give evidence
of non-linear coupling between spiral waves, with an even stronger
efficiency than the example presented here.}

This paper is organized as follows~: in a first part we present a
general background about mode coupling.  We present the selection
rules and justify the high efficiency of coupling when the frequencies
of the coupled waves are such that their resonances coincide.  Since
the physics of coupling can be understood without heavy mathematical
derivations, we have avoided lengthy and intricate details
on the derivation of the coupling efficiency, which can be found in
the references given in Sygnet \etal 1988, or in Masset \& Tagger
1996b.  In a second part
we present the characteristics of the code, and in the third part we
present the results of a run which shows the unambiguous signature of
non-linear coupling between the bar and $m=0$, $m=2$ and $m=4$ spiral
waves.  An  additional run without the central bar is also presented
in order to show that non-linear coupling is indeed responsible for
the behavior observed in the external parts of the galaxy (\ie the
excitation of a slower spiral whose ILR coincides with the
corotation of the bar).

\section{General notions on non-linear coupling}

\subsection{Notations}
We will note $m$ the azimuthal wavenumber of a wave, which
is an integer corresponding to the number of arms of this wave.
We note $\omega$ the frequency of a wave in the galactocentric
frame, with $\Omega_p=\omega/m$ the pattern frequency (in the
following, including the plots, we will primarily label waves by
$\omega$ rather than $\Omega_{p}$).

Finally we note $\Omega(r)$ the angular rotation velocity of the stars,
$\kappa(r)$ the epicyclic frequency, $\sigma_r$ and $\sigma_\theta$ the
radial and azimuthal
velocity dispersion.

\subsection{Non-linear coupling and selection rules}

\label{sec:selrule}

Mode coupling is a very specific non-linear mechanism (see \eg
Laval \& Pellat 1972 and Davidson 1972 for a general discussion).  It contrasts
with the usual picture of strong turbulence, where a large number of
modes interact, forming in the asymptotic limit a turbulent cascade
(\eg the Kolmogorov cascade in incompressible hydrodynamics).  This
asymptotic limit is reached when a whole spectrum of waves or modes is
excited, with a very small correlation time, so that each mode exists
only for a very short time before it looses its energy to others (\eg
in the Kolmogorov spectrum the correlation time is of the order of the
eddy turnover time).  Mode coupling, on the other hand, occurs in
situations where only a small number of waves or modes can exist, so
that each interacts non-linearly with only a few others -ideally only
two.  In particular we will see in our numerical results that if the
two $m=2$ patterns interact non-linearly with an $m=4$, the $m=6$ that
results from the coupling of one $m=2$ with the $m=4$ can be clearly
identified, but remains so weak that its influence can be neglected
(technically, since the $m=4$ results from the non-linear interaction
of two waves, it is associated with quadratic terms in the
hydrodynamical equations; then the $m=6$, resulting from the coupling
of the $m=4$ with one $m=2$, is associated with third-order
terms, which remain small).  This small number of active modes
translates into long correlation times, \ie the quasi-stationary
structure found in the simulations.

In a linear analysis all the waves present in the disk behave
independently and do not interact.  If one retains higher order terms
of the hydrodynamical or kinetic equations, this is no more true and
waves can exchange energy and angular momentum, provided that they
fulfill a number of {\it selection rules}.

Let us consider two spiral density waves~1 and~2,
with azimuthal wavenumbers
$m_1$ and $m_2$ and frequencies in the galactocentric frame $\omega_1$
and $\omega_2$. The perturbed quantities relative to each wave will
be of the form~:

\[\xi_{1,2} \propto e^{i(\omega_{1,2}t-m_{1,2}\theta)}\]

where $\xi$ stands for one of the perturbed quantities (a velocity
component, the density or the potential).  In the hydrodynamical or
kinetic equations, one finds cross products of the form $\xi_1\xi_2$
and $\xi_1\xi_2^*$ (where the $*$ notes the complex conjugate), arising
from non-linear terms (\eg the \(\vec v \cdot \vec \nabla \vec v\) or
the $\rho\vec\nabla\Phi$ terms of the Euler equation).  These terms
correspond to the beat waves associated with the perturbed quantity~:

\[\xi_B \propto e^{i[(\omega_1\pm\omega_2)t-(m_1\pm m_2)\theta]}\]

Hence the beat waves will have the frequency and wavenumber~:

\begin{equation}
\label{eqn:selrule_m}
m_B = m_1\pm m_2
\end{equation}
and

\begin{equation}
\label{eqn:selrule_o}
\omega_B = \omega_1\pm \omega_2
\end{equation}

Thus when one performs a Fourier analysis over time and azimuthal
angle of the Euler equation, at the frequency $\omega_{B}$ and the
wavenumber $m_{B}$ one will find both {\it linear} terms directly
proportional to the amplitude of the beat wave, and {\it quadratic}
ones proportional to the product of the amplitudes of waves 1 and 2.
These terms act as a sink or source of energy for the beat wave.  Two
cases can then occur: the frequency and wavenumber of the beat wave
may correspond to a perturbation which cannot propagate in the system;
this perturbation is thus simply forced by the ``parent'' waves.  But
they may also correspond to a wave which can propagate (\ie they obey
the linear dispersion relation); this means that the system can
spontaneously oscillate at the frequency and wavenumber excited by the
parents and will thus respond strongly to the excitation: just as a
{\it resonantly driven} oscillator, the beat wave can reach a large
amplitude\footnote{A small point must be made about the vocabulary:
because of this analogy with a resonant oscillator, this process is often
called {\it resonant} mode coupling in the relevant litterature.  Thus
this term is independent of the additional factor, discussed below, that
in our case the coupling occurs at a radius where the waves involved
have a resonance with the particles.}.  We will see below that this is
the case we study.

In that situation, since the three waves obey the linear dispersion
relation and can reach large amplitudes, and since each corresponds to the
beating of the other two, they can no more be distinguished as ``parent''
and ``beat'' waves: one has a system where the three waves play similar
roles and can strongly interact, exchanging their energy and angular
momentum.  In the homogeneous (and thus much more simple) physical systems
where this has been well studied, it has been found to result either in
stationary behaviors (\eg one wave is linearly unstable and extracts free
energy from the system; the other two are linearly damped and can saturate
the growth of the first one at a finite amplitude, by dumping the energy in
a different form), or in cyclic (the classical Manley-Rowe cycles) or even
chaotic behaviors (see \eg Laval \& Pellat 1972, Davidson 1972, Meunier
\etal 1982).

For a set of three waves such that their frequencies and azimuthal
wavenumbers fulfill the relations (1-2), which are the {\it selection
rules}, one finds linearly that the energy and angular momentum fluxes
that each carry are constant; non-linearly, one finds that the time
derivative of the
energy density of each wave is proportional to the product of the
amplitudes of the other two, \ie these waves exchange energy and angular
momentum.  The derivation of the exchange rate can be done using the
formalism of quadratic variational forms, and is beyond the scope of this
paper devoted mainly to numerical results.  The derivation of the
efficiency of the non-linear coupling of bar and spiral waves is given in
kinetic formalism by Tagger \etal 1987 and Sygnet \etal 1988, and in
hydrodynamical formalism for the coupling of spiral and warp waves by
Masset \& Tagger, 1996b.

These papers also explain why we have not introduced in the above
discussion the radial and vertical wavenumbers of the waves, which
should play {\it a priori} the same role as the azimuthal wavenumbers
or the frequencies in the selection rules.  The reason is that, in
theses directions, the system is inhomogeneous so that the Fourier
decomposition of linear waves is irrelevant.  In the vertical
direction, the waves have a standing structure and one finds that the
coupling coefficient is simply proportional to a scalar product of the
vertical structure functions (the Fourier integrals performed in the
azimuthal direction are just a particular case of this scalar
product).  In the radial direction, we will find that the coupling
occurs over a very narrow annulus, so that the waves essentially
``ignore'' their radial wavelengths (which might be derived in a WKB
approximation).  The next section explains why we expect coupling to
occur efficiently over a small radial extent.  Thus, for a wave which
for instance receives energy by non-linear coupling, one can consider
that it is excited at that very precise radius, with a frequency and
azimuthal number given by the selection rules.  It will then propagate
radially with a radial wavenumber given by its local dispersion
relation, independent of the radial wavenumbers of the other waves.

\subsection{Localization of the coupling}

\label{sec:localization}

As mentioned in the preceding section, we can write the time
derivative of the energy density of each wave as a coupling
term involving the product of amplitudes of the other two.  This
time derivative is performed following the wave propagation, \ie~:
\[dE/dt \equiv \partial E/\partial t + (1/r)\partial(rc_gE)/\partial r =
\mbox{Coupling Term}\]
where $c_g$ is the group velocity, \ie the
velocity at which energy is advected radially, and the right-hand side
vanishes in linear theory.

\begin{figure}
\psfig{file=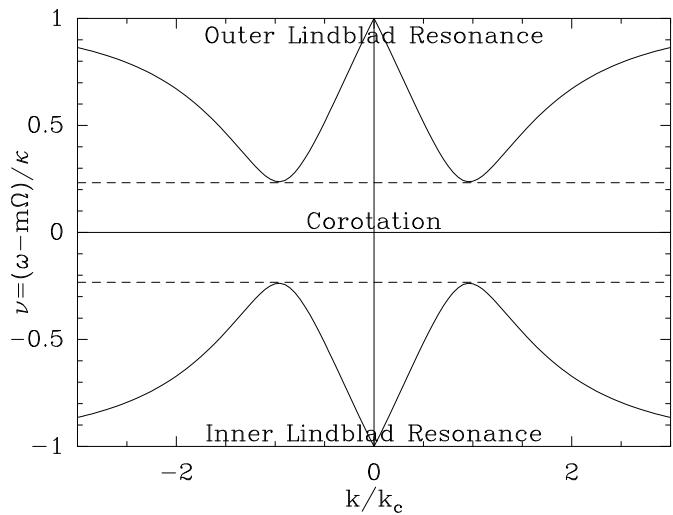,width=\columnwidth}
\caption{\label{fig:rdstellar}
This figure shows the dispersion relation of a stellar spiral density wave,
in the
WKB (\ie tightly wound) approximation, for a disk with a Toomre
parameter $Q=1.05$.  The critical wavevector $k_c$ is given by~: $k_c
= \kappa/\sigma_r$.  The quantity $\tilde\omega=\omega-m\Omega(r)$ is
the wave frequency measured in the frame rotating with the stars. It
vanishes at corotation and is equal to the epicyclic frequency at the
Lindblad Resonances.  The group velocity $c_g =
\partial\tilde\omega/\partial k$ vanishes when the wave reaches the
forbidden band around corotation (delimited by dotted lines) or the
Lindblad Resonances, characterized by $\nu =\tilde\omega/\kappa= \pm
1$ and $|k/k_c| \gg 1$.  Note that at small $k/k_r$, the wave is not
tightly wound and the WKB approximation fails, so that the dispersion
relation cannot be used for $|k/k_c| \ll 1$.}
\end{figure}

If we assume for simplicity a stationary state (we will see from our
numerical simulations that this assumption is not too far from reality), we
can write~:

\[\partial(c_gE)/\partial r = \mbox{Coupling Term}\]

where we have neglected the effect of cylindrical geometry, since we
expect the coupling to be very
localized
radially\footnote{This assumption will be checked {\it a posteriori}
on the numerical results in section~4.3;
we will see that even when the coupling partners coexist on a wide
range of radii, the coupling efficiency (\ie the energy and angular
momentum exchange between these modes) is strongly peaked on a narrow annulus
which will be identified as the
corotation of the bar.}. This shows in
particular that wherever $c_g$ vanishes, the variations of $E$ can be
strong even if the coupling term is not large.  In fact $c_g$ vanishes at
Lindblad resonances and at the edge of the forbidden band around
corotation (Toomre 1969), as can be seen from
Fig.~\ref{fig:rdstellar}, so that we can expect non-linear coupling
to be highly efficient when the waves are close to these radii.  The
physical meaning of this enhanced efficiency is simply that the waves
stay in these regions for a long time, so that they can be efficiently
driven, and exchange a sizable fraction of their energy, even at low
energy transfer rates.

Let us note finally that if one of the waves is at its Lindblad resonance
and a second one at its corotation, their beat waves, because of the
selection rule on frequencies, are also at a Lindblad resonance, still
improving the efficiency of non-linear coupling.  This appears in a
different form in the kinetic description used by Tagger \etal 1987 and
Sygnet \etal 1988: there the coupling coefficients appear as integrals over
phase space of the stellar distribution, containing two resonant
denominators of the form $\omega-m\Omega$ and $\omega-m\Omega\pm\kappa$
(instead of classically only one in the linear terms leading to Landau
damping), thus making the coupling efficiency very efficient when the
resonances of the waves coincide.

\section{Numerical implementation}

\subsection{Algorithms}
We have written a classical Particle-Mesh (PM) two-dimensional code
simulating the stellar component of disk galaxies, with special emphasis on
diagnostics adapted to the physics we describe.  We have not taken into
account the gaseous component, which is not expected to modify dramatically
the coupling mechanism. A more detailed discussion about the role a
dissipative component could play will be given in section~4.4.2.

The density is tabulated on a cartesian grid using the bilinear
interpolation {\it
Cloud in Cell} (CIC).  The potential is computed using a FFT algorithm with
a doubling-up of the grid size in order to suppress tidal effects of
aliases (see Hockney \& Eastwood 1981).
We use a softened gravity kernel ($\sim
-G/(r^2+\epsilon^2)^{1/2}$) for the computation of the potential, with a
softening parameter $\epsilon$ chosen so as to mimic the effect of the disk
thickness.  Only the stars which are in the largest disk included in the
grid are taken into account for the evaluation of the potential, in order
not to artificially trigger $m=4$ perturbations.

The force on each star is computed from the potential using a CIC scheme,
so that the stars undergo no self-forces. The positions and velocities are
advanced using a time-centered leap-frog algorithm.

Finally we have added three unresponsive static components~: a
central mass,
a bulge and a halo. The analytical force law corresponding to each
of these components is given in table~\ref{tab:comp}.

\begin{table}
\begin{center}
\begin{tabular}{|c|l|l|}
\hline
Component \phantom{\Large{M}} & \multicolumn{1}{c|}{Parameters} &
\multicolumn{1}{c|}{Central acceleration} \\
\hline \hline
Central Mass  	& mass $M_c$  			& \enlarge $\dss
\frac{GM_c}{r^2+\epsilon^2}$\\
\hline
Bulge		& mass $M_b$			& \enlarge $\dss
\frac{GM_br}{a(b+a)^2}$\\
		& radius $b$			& where $a = \sqrt{b^2+r^2}$\\
\hline
Halo		& limit speed $v_\infty$ 	& \enlarge $\dss
\frac{r_cv_\infty^2}{r^2}{\cal L}(r/r_c)$\\
		& core radius $r_c$		&  ${\cal L}(x) =
x-\tan^{-1} x$\\
\hline
\end{tabular}
\end{center}

\caption{\label{tab:comp}
This table shows the analytical law for the central acceleration of the
three static components, as well as the characteristic parameters defining
each of them.  The halo is not characterized by its mass which would be
infinite if it extended to infinity. It is characterized instead by the
asymptotic rotational velocity reached far from the galactic center.}
\end{table}

\subsection{Initialization}

At the simulation startup, particles are placed at random radii
with a probability law resulting in an exponential surface density
profile:
$\Sigma(r) = \Sigma_0e^{-r/R_d}$, where $\Sigma_0$ is the central surface
density and $R_d$ is the length-scale of the galactic disk.  All the
particles have the same mass.  The velocities are randomly assigned using
the epicyclic approximation so that~:

\begin{itemize}
\item{The velocity distribution be a local anisotropic Maxwellian.}
\item{The ratio of the radial to azimuthal dispersion,
$\sigma_r/\sigma_\theta$ be $2\Omega/\kappa$, with $\Omega$ and $\kappa$
computed consistently from the static and stellar potentials.}
\item{The average azimuthal velocity be the rotational velocity
corrected by the Jeans drift arising from the gradient of surface
density.}
\item{The Toomre $Q$ parameter be constant over the whole disk.}
\end{itemize}

Since the epicyclic approximation fails close to the galactic center,
these
prescriptions do not result in an exact equilibrium distribution in the
central regions; we thus see in the first dynamical times of our runs
relaxation oscillations, leading to a slight radial redistribution of
matter in the vicinity of the galactic center.

The truncation radius of the disk corresponds to the edge of the
active grid. This grid is chosen large enough so as to ensure that
the runs are not perturbed by edge modes. For instance in the runs
we present below the average number of particles per cell in the
most external cells of the active part of the grid is $6.5\cdot 10^{-3}$,
\ie totally negligible.

\subsection{Tests}

The behavior of the code has been tested so as to ensure that~:

\begin{itemize}
\item{a single massive particle follows Newton's first law, \ie is not
subject to self-force;}
\item{two particles obey the 2-bodies laws within errors arising
from finite cell size and finite timestep;}
\end{itemize}

Furthermore, during a run, we check that the total angular momentum is
exactly conserved (within the numerical precision errors), and that the
total energy is properly conserved (within 10~\%) over the whole duration
of the run ; we monitor the fraction of stars ejected out of the active
grid, so as to be sure that it remains sufficiently small.

\subsection{Spectral analysis}

The purpose of our numerical simulations is to check for the presence
of non-linear coupling between waves and modes (simply defined here as
quasi-stationary structures: they can thus be either linear
eigenmodes, if they stay at low amplitude and unaffected by non-linear
effects, or more complex non-linear entities as will be found below).
We do this by identifying the features arising during the run, and
checking the frequency and wavenumber relations between them.

For this we plot spectral density contours, in the same manner as
Sellwood 1985 : at each output time we perform a Fourier transform of the
perturbed density in the azimuthal direction, resulting for each value of
the azimuthal wavenumber $m$ in spectra depending on time and radius. In
order to avoid grid artifacts, we compute the spectra directly from the
coordinates of each particle rather than by a Fourier Transform of the
interpolated grid.  Hence for each value of $m$, we compute~:
\[W_C^{(m)}(r,t) = \sum_{i=1}^{N}\cos(m\theta_i)b(r,r_i)\]
and
\[W_S^{(m)}(r,t) = \sum_{i=1}^{N}\sin(m\theta_i)b(r,r_i)\]

where $(r_i,\theta_i)$ are the polar coordinates of the $i^{th}$
particle, $N$ is the total number of particles, and,
for a given $r_i$, $b(r,r_i)$ is a
``bin''-function which linearly interpolates the value on a
monodimensional radial grid.

The temporal spectrum is obtained by taking the Fourier Transform
$\tW^{(m)}(r,\omega)$ with respect to time of the complex function
$W_C^{(m)}(r,t)+i W_S^{(m)}(r,t)$.  We then plot either the amplitude or
the power spectrum (\ie either $|\tW^{(m)}(r,t)|$ or $|\tW^{(m)}(r,t)|^2$)
properly normalized so as to represent either the relative perturbed
density in the case of the amplitude spectrum, or the energy density in the
case of the power spectrum.

Unlike Sellwood (1985), we eliminate the first timesteps when computing
the temporal Fourier transform. Indeed these first timesteps correspond
to a transient regime, and taking them into account degrades the spectrum
and makes its interpretation more complex.
The choice of the first timestep used to compute the Fourier transform
is made by looking at the $W_{C,S}^{(m)}(r,t)$~plots. We start the Fourier
transform when the features observed on these plots appear to have
settled to the quasi-periodic behavior always obtained after a few
rotation periods.

\section{Results}

We present two complementary runs.  The first one exhibits a simple,
typical example where a strong bar develops, together with a slower
outer spiral; we confirm that their frequencies are such that the corotation
of the bar coincides with the ILR of the spiral.  The second run is
performed with the same parameters, except that we inhibit the bar by
changing the rotation profile at the center, so that the bar gets
damped at its ILR. The rotation profile in the outer parts is not
changed, and we check that the outer spiral obtained in the first run
does not develop in this second one: this proves that its formation
was not due to local conditions, but indeed to the non-linear excitation by
the bar.

\subsection{Run 1}
In this run we use a $128\times 128$-active mesh with $600$~pc wide cells.
The
galaxy is an exponential disk of 80,000 particles with total mass
$6.1\cdot 10^{10}$~$M_\odot$ and a characteristic length $R_d=3.5$~kpc.  The
softening length $\epsilon$ is $300$~pc.  The Toomre $Q$ parameter
is initially constrained to be $1.3$ over the whole disk.  The disk is
embedded in a static halo and a static bulge.  The halo has a core radius
$r_c=2$~kpc and an asymptotic speed $v_\infty=120$~km/s, so that the mass
inside the smallest sphere containing the whole disk is $1.2$ that of the
disk.  The bulge has a radius $b=2$~kpc and its mass is $M_b=5\cdot
10^{10}$~$M_\odot$.
There is no central point-like mass, \ie $M_c = 0$.
We use a timestep of $0.75$~Myr, we perform the
simulation over 16,000 timesteps, and we output the grid density,
$W_{C,S}^{(0,1,2,3,4)}(r,t)$ and some other quantities (velocity
dispersion, energy, etc.) every 20~timesteps.

The chosen grid is oversized for the study of the disk, in order to
avoid edge effects which have appeared to modify strongly the behavior
of previous runs.  Thus in all the plots and spectra we present in
this paper, we have eliminated the outer parts of the active grid,
where results are quite noisy due to the rarefaction of stars.  In
particular, all the spectra are presented on the range 0-27~kpc.

\begin{figure}
{
\parbox{4cm}{
\psfig{file=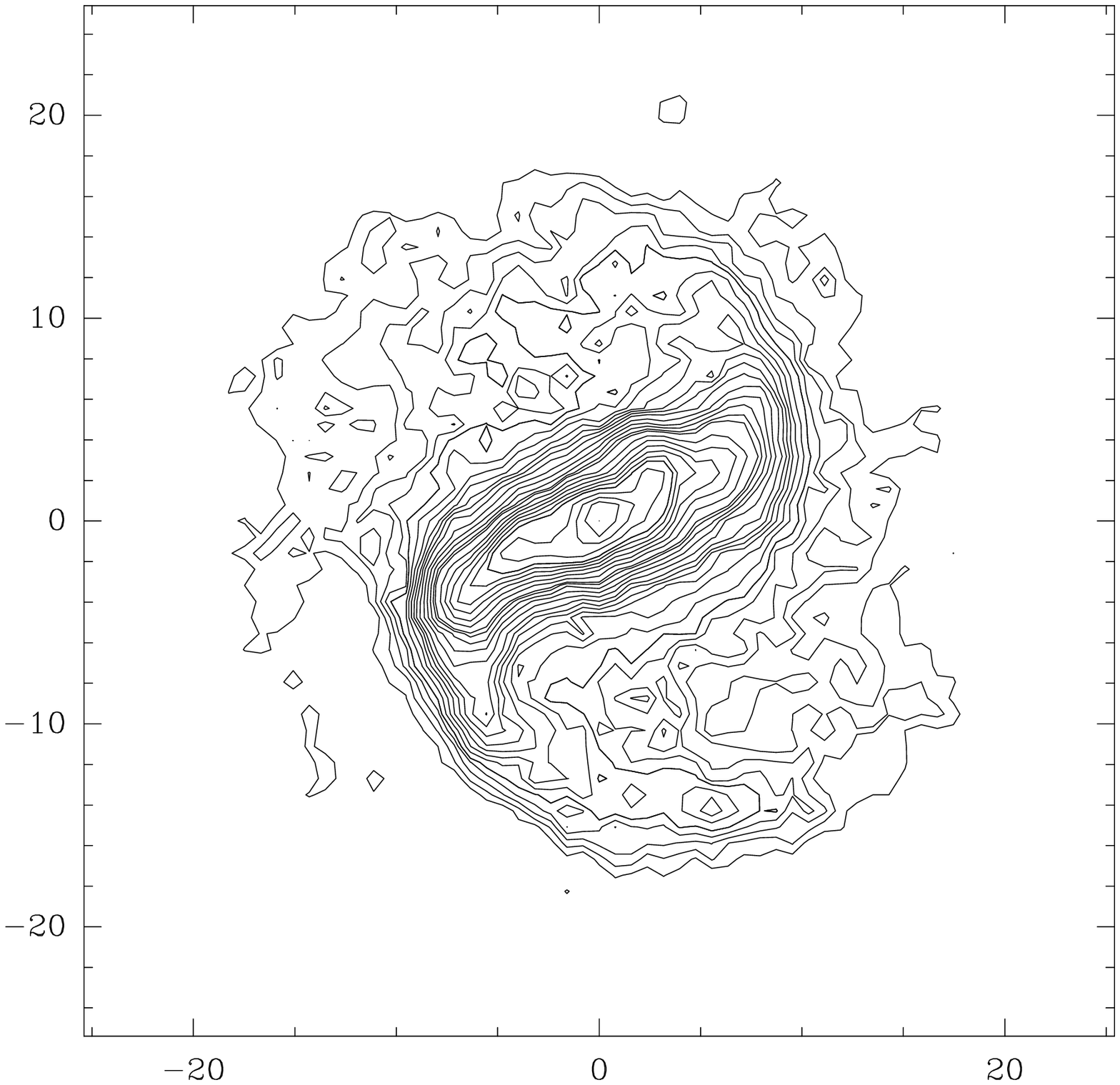,width=4cm}
} \hfill
\parbox{4cm}{
\psfig{file=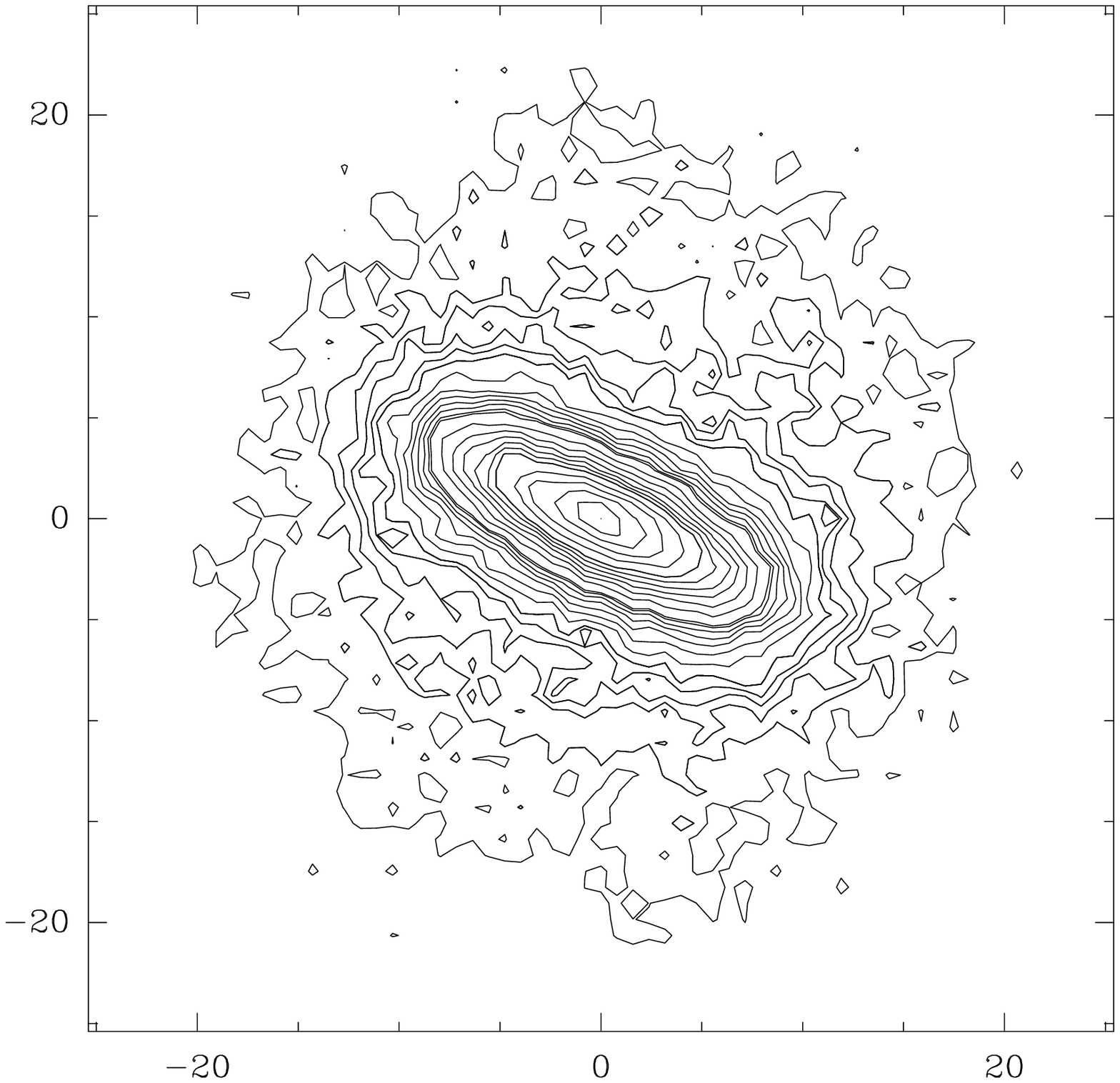,width=4cm}
}}
\caption{\label{fig:run1_pictures}
The left plot is an isocontour of the grid density of run~1
at time~900~Myr. The
right one represents the same quantity at time~9000~Myr. The coordinates
scale is in kpc.
}
\end{figure}

The galaxy develops a strong bar which appears a bit before~1~Gyr, and
triggers a strong spiral wave outside corotation.  The bar and the
spiral heat the disk so that the spiral arms weaken due to the
decreasing efficiency of the Swing mechanism with increasing Q.
Fig.~\ref{fig:run1_pictures} shows two plots of the particle
density, the first one when the bar appears, and the second one
close to the end of the run, when the disk is quite hot.  No striking
spiral feature appears on this last plot in the outer part of the
galaxy.

The simulation could be made more realistic if dissipation were included
through the presence of a gaseous component, and taking into account star
formation which would continuously cool the stellar population and maintain the
efficiency of the Swing mechanism, so that we would still have a noticeable
spiral structure even at late times.  Also, since the response of the
gas is strongly non-linear, we expect that it would make the outer spiral
structure more prominent. But here we focus on the
non-linear coupling, so that the absence of dissipation is not
critical for our purpose.  We will discuss later the influence it should
have on the coupling.

\begin{figure}
\psfig{file=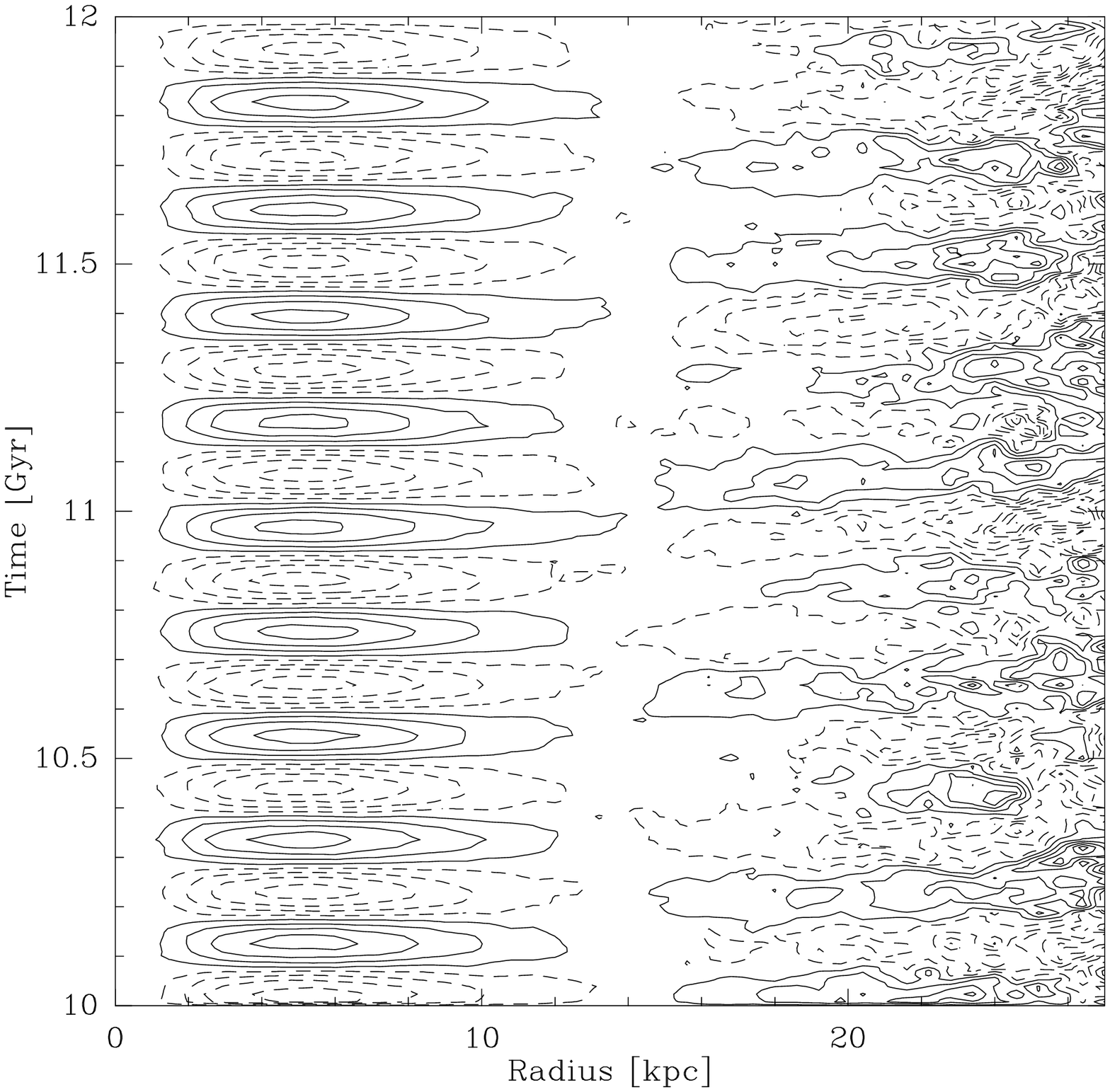,width=\columnwidth}
\psfig{file=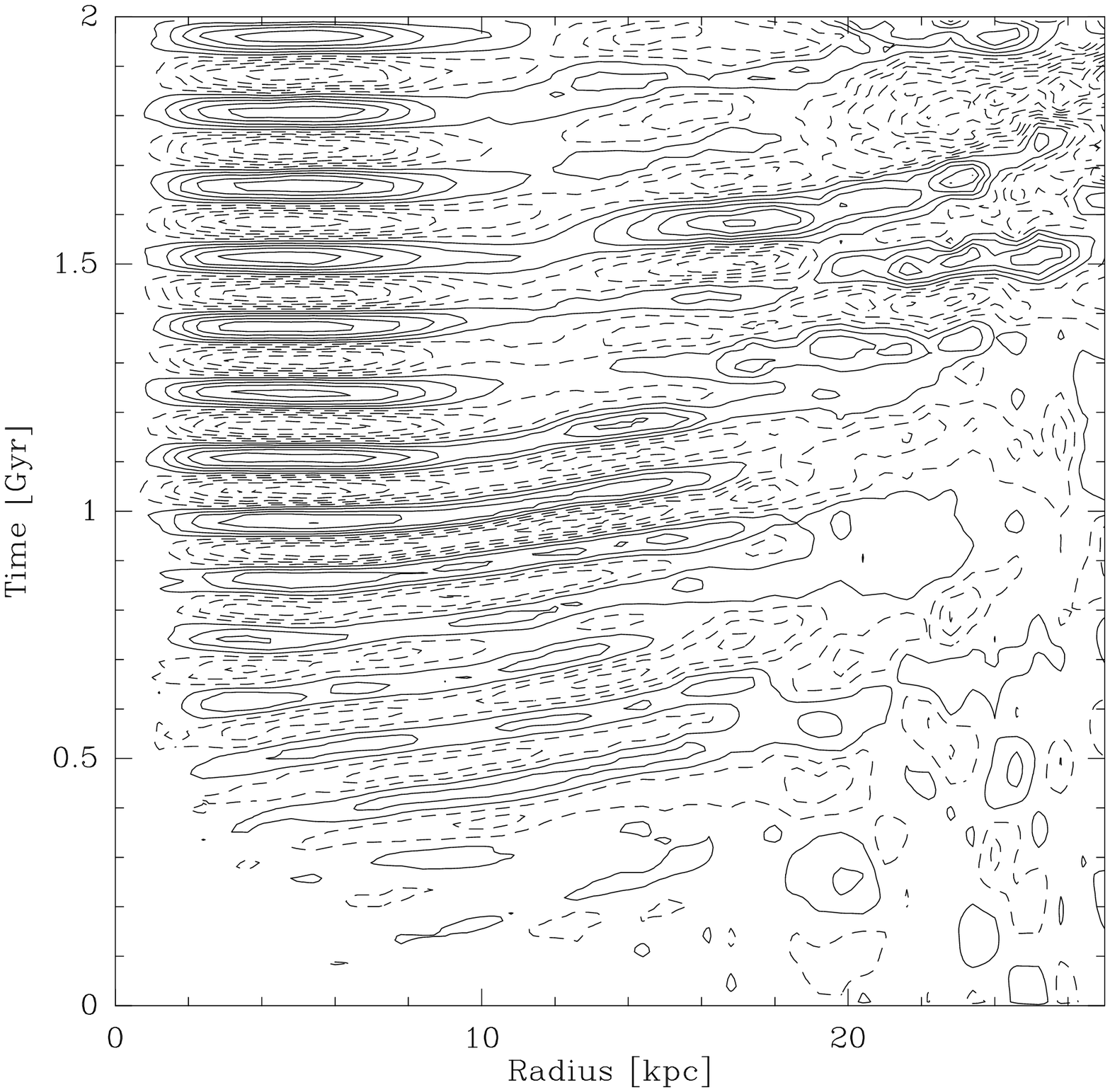,width=\columnwidth}
\caption{ \label{fig:run1_trm2}
The lower plot is an isocontour representation of $W_C^{(2)}(r,t)$
normalized so as to represent the relative perturbed density,
with time varying between $0$~and $2$~Gyr. The upper plot
represents the same function with time between $10$~Gyr and
$12$~Gyr.
}
\end{figure}

Let us now turn to the power spectra of the perturbed density.  On
Fig.~\ref{fig:run1_trm2} we have plotted isocontours of the function
$W_C^{(2)}(r,t)$, \ie the cosine contribution of $m=2$ density
perturbations as a function of radius (abscissa) and time (ordinate).
This can be seen as the point of view of an observer located at radius
$r$ and $\vartheta = 0$, and seeing the bar and spiral arms sweep by
with time.  The lower plot represents the beginning of the simulation,
from the initial time step to $2000$~Myr, while the upper one shows
the very end of the simulation, with time varying from $10$~Gyr to
$12$~Gyr.

\begin{figure}
\psfig{file=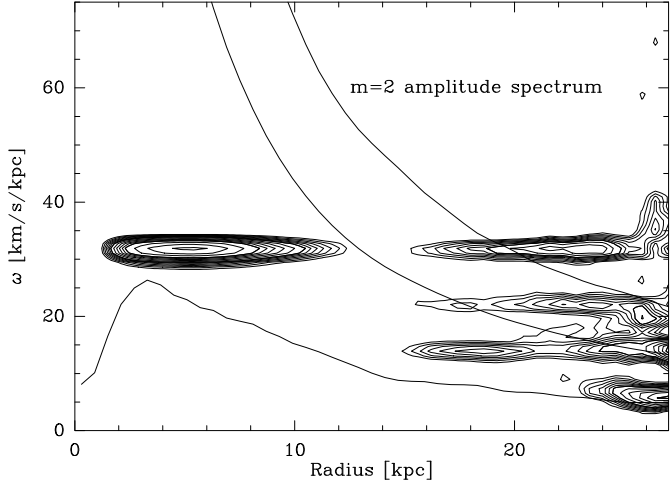,width=\columnwidth}
\caption{\label{fig:run1_power_m2}
This figure shows the $m=2$ amplitude spectrum of the relative
perturbed density. The three curves show respectively
$2\Omega-\kappa$ (which gives the location of the ILR of $m=2$ waves,
depending on their frequency), $2\Omega$
(which gives the corotation) and $2\Omega+\kappa$ (which
gives the OLR) computed at time $t=8145~Myr$, \ie close to the
middle of the time interval over which the Fourier Transform
is performed.
The contour spacing is $10^{-2}$, and the first contour
level is $4\cdot 10^{-2}$.
}
\end{figure}

On the lower (earlier) plot we first see transient features which
propagate outward, as shown by their obliquity.  Between 1~Gyr and
1.5~Gyr we see the bar form, resulting in purely horizontal (\ie
standing) features.  The plot clearly shows that the bar forms a
quasi-stationary structure ending about 10~kpc from the galactic
center, \ie in the region of its corotation.
Its corotation will slowly move
outward as the bar adiabatically slows down at later times, as often
seen in this type of simulations
(Pfenniger \& Friedli~1991, Little \& Carlberg~1991).
As expected from linear mode
theory, the bar extends beyond corotation as a spiral wave propagating
outward.  The upper (later) plot shows how robust the bar is, since it
has lasted for about 10~Gyr with a frequency that has slowly decreased.
The right part of the diagram shows that there are still features
propagating outside corotation, but that they are much less regular
than the spiral waves formed from the ``young'' bar (in the upper part
of the lower plot).  This behavior which might be believed chaotic is
in fact easily understood from the time
Fourier Transform of this diagram (also taking into account
$W_S^{(2)}(r,t)$ as imaginary part), \ie the $m=2$ spectral density,
shown on Fig.~\ref{fig:run1_power_m2}. The Fourier transform is
performed over~512 outputs (from~288 to~799, \ie for time ranging from
$4320$~Myr to $11985$~Myr).

\begin{figure}
\psfig{file=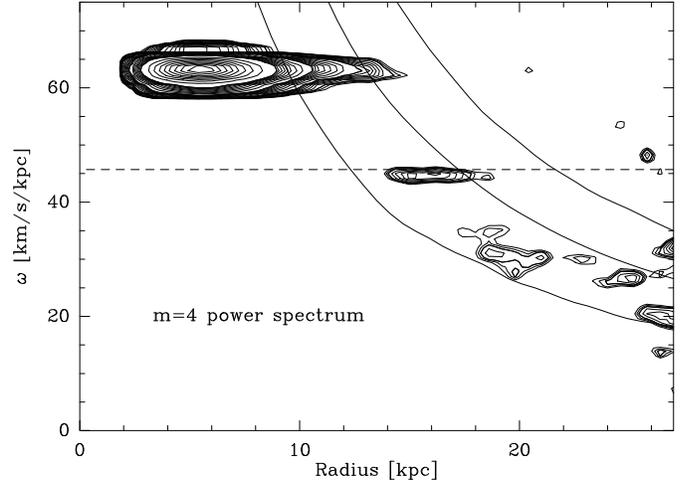,width=\columnwidth}
\caption{\label{fig:run1_power_m4}
This figure shows the $m=4$ spectrum of the energy
density. The three curves are respectively
$4\Omega-\kappa$ (which gives the ILR of $m=4$ waves), $4\Omega$
(which gives the corotation) and $4\Omega+\kappa$ (which
gives the OLR) computed as in figure~\ref{fig:run1_power_m2} at
time $t=8145$~Myr.
Since the radial normalization factor includes a term in $\Sigma(r)$,
which has an exponential behavior, the bar is strongly enhanced and
the contour levels must be adjusted in order to reveal the expected $m=4$
beat wave. This explains the behavior of the successive isocontours
in the bar harmonic at $\omega=64$~km/s/kpc.
When normalized so as to represent the amplitude of perturbed
density, this $m=4$ spectrum gives a relative perturbed density
$\sigma/\Sigma \simeq 2\cdot 10^{-2}$ for the beat wave.
Taking the energy density power spectrum was necessary here to avoid
strong noisy contributions on the right of diagram due to particle
rarefaction at the outer edge of the disk. The dashed line indicates
the expected frequency of the $m=4$ coupling partner, \ie $\omega_B
+\omega_S$.
}
\end{figure}

\begin{figure}
\psfig{file=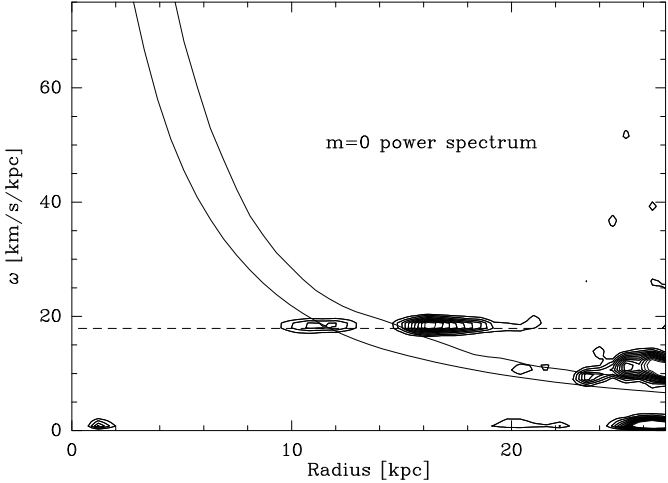,width=\columnwidth}
\caption{\label{fig:run1_power_m0}
This figure shows the $m=0$ power spectrum of the relative perturbed
density.  The two curves are respectively $\Omega$ and $\kappa$
at time $t = 8145$~Myr.
Taking the energy density power spectrum was once again necessary
here to avoid
strong noisy contributions on the right of diagram. The dashed line
indicates the expected frequency of the $m=0$ coupling partner, \ie
$\omega_B-\omega_S$.
}
\end{figure}

The thinness of the features obtained shows that they correspond to
quasi-stationary structures (bar or spiral) in the disk.  In the inner
part we see the strong contribution of the bar, which stops at
corotation.  Outside corotation we have two structures at different
frequencies, explaining the apparent lack of periodicity in this
region in Fig.~3 (there is also an intermediate, weaker structure at
$\omega\simeq 22$~km/s/kpc, for which a tentative explanation will be
given later).  The faster wave has the same frequency as the bar and
corresponds to the spiral wave excited by the bar through the Swing
mechanism.  The slower one appears to have its ILR at approximately
the same radius as the corotation of the bar, as expected from the
works of Tagger \etal 1987 and Sygnet \etal 1988.  In order to check
that this second mode is fed by non-linear coupling, we have to check
for the presence of a beat wave near the bar corotation radius.  Let
us call $\omega_B$ the bar frequency and $\omega_S$ the frequency of
the second (lower) spiral.  We measure the frequencies from the maxima
on the isocontours, with a typical accuracy $\pm 0.5$~km/s/kpc.  We
find $\omega_B = 31.8$~km/s/kpc and $\omega_S = 13.9$~km/s/kpc.  Hence
according to equations (\ref{eqn:selrule_m}) and (\ref{eqn:selrule_o})
we have to check on the $m=2+2$ spectrum for the presence of a mode at
frequency $\omega_B+\omega_S=45.7$~km/s/kpc, and on the $m=2-2$
spectrum (\ie an $m=0$ mode which would appear as a ring in the
structure of the galaxy) for the presence of a mode at frequency
$\omega_B-\omega_S=17.9$~km/s/kpc.  These spectra are presented
respectively on Fig.~\ref{fig:run1_power_m4}
and~\ref{fig:run1_power_m0}, and the expected frequencies are
indicated by a dashed line.

We have chosen to represent all the spectra of  the run with the
same coordinates scale. This enables the reader to check graphically
for the selection rules by superimposing copies of the spectra.

On the $m=4$ spectrum we see a major contribution corresponding to the
first harmonic of the bar at $2\omega_{B}$, and a weaker one which is at the
expected frequency, and which begins close to its ILR, \ie also close to
the expected coupling region.

Similarly on the $m=0$ spectrum we see that a major contribution comes
from the expected beat wave, which is once again located close to the
corotation of the bar.  More precisely, the measured frequencies are
$\omega_4 = 44.7$~km/s/kpc and $\omega_0 = 18.3$~km/s/kpc, in
excellent agreement with the expected ones (within $2$~\% for both
waves).

However, the discussion above is not sufficient to ensure that the
slower spiral is non-linearly excited at the bar corotation.
Two questions still remain:
\begin{itemize}
\item{Is the slower spiral really triggered by the bar, or does it
exist independently of it~? The presence of $m=0$ and $m=4$ beat
waves would still be expected in such a case, since we are dealing
with finite amplitude waves, but only as ``passive'' features
proportional to the product of the amplitudes of the parent $m=2$
waves.}
\item{Is the coupling localized in the bar corotation/slower spiral ILR
region, as expected from the theoretical work of Tagger \etal 1987 and
Sygnet \etal 1988~?}
\end{itemize}

In order to answer the first question, we have performed a second run,
almost identical to the first one, but where we have inhibited the bar
formation.

\subsection{Run 2}
\begin{figure}
\psfig{file=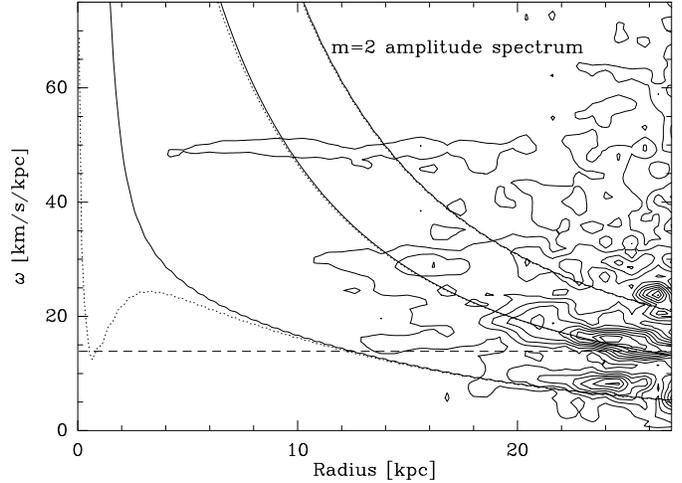,width=\columnwidth}
\caption{\label{fig:run2m2}
On this figure we show the m=2 amplitude spectrum of run~2. The
contour spacing is $10^{-2}$, and the first contour is at $10^{-2}$.
The solid lines represent the ILR, corotation and OLR curves for run~2,
and the dotted lines the same quantities for run~1. The dashed horizontal
line represents the frequency at which the slower spiral was observed
in run~1.}
\end{figure}

In this run we have taken a bulge mass of $M_b=4.3\cdot 10^{10}\,M_\odot$
and a central point-like mass of $M_c = 7\cdot 10^9\,M_\odot$. All the other
parameters are the same as in run~1. The sum of
the bulge mass and the central mass equals the bulge mass in run~1,
so that the rotation curve
(and thus all the characteristic frequencies) coincide with the ones of
run~1 at radii larger than the bulge radius. The only difference is
that in this new run the ILR curve has no maximum, and thus prevents
the formation of a stellar bar (or, to put it differently, the central mass
is 11.4\% that of the stellar disk, far above the critical value of
about 5\% which is thought to be sufficient to destroy the bar,
(see \eg Norman \etal 1996), or even lower (2--3\%, see
Friedli~1994).
The $m=2$ amplitude spectrum, computed in exactly the same conditions
as in run~1, is presented in Fig.~\ref{fig:run2m2}.

Some faint standing modes (\ie thin features) appear on this figure,
but they all are far fainter than the bar and spirals of run~1
(except maybe at the outer edge where particle noise may become
large).  In particular one should note that the first contour level
for Fig.~\ref{fig:run1_power_m2} was $4\cdot 10^{-2}$, whereas it is
$10^{-2}$ in Fig.~\ref{fig:run2m2}.  Obviously, in absence of the
central bar, there is no standing
mode in this run~2 at the frequency of the slower spiral in run ~1,
and the mode which appears close to this frequency is far fainter than
the slow spiral in run~1.

\begin{figure}
\psfig{file=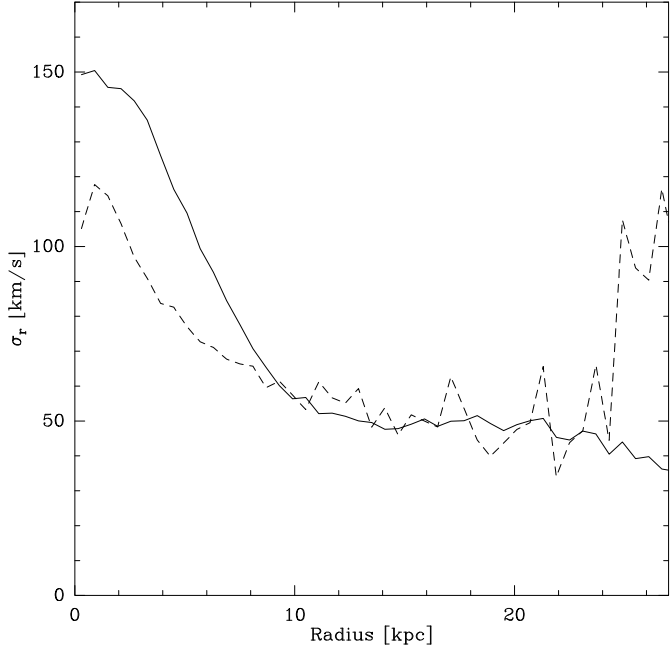,width=\columnwidth}
\caption{\label{fig:compsigma}
Comparison of radial velocity dispersions at time $8175$~Myr for run~1
(solid line) and run~2 (dashed line).  The disk run~1 has been heated
in its central region by the bar, but its temperature is similar to
that of run~2 at radii larger than $\sim 10$~kpc, except for a
strong peak in run~2 at the outer edge, which could not damp a wave
but possibly reflect it by acting as a Q-barrier.}
\end{figure}

The comparison of the stellar dispersion velocity between run~1 and
run~2 is presented on  Fig.~\ref{fig:compsigma}. In the radial range
where the slower spiral existed in run~1, we see that the disk
temperature is the same for run~1 and run~2.

Hence run~1 and run~2 disk are very similar (orbital and epicyclic
frequencies, temperature, density) in the region over which the slower
spiral of mode~1 extended, and nevertheless no such mode appears
in run~2.  This is a strong point in favor of our justification of
this slow spiral as non-linearly triggered by the bar.

\subsection{Radial behavior of wave amplitudes in Run~1}
\begin{figure}
\psfig{file=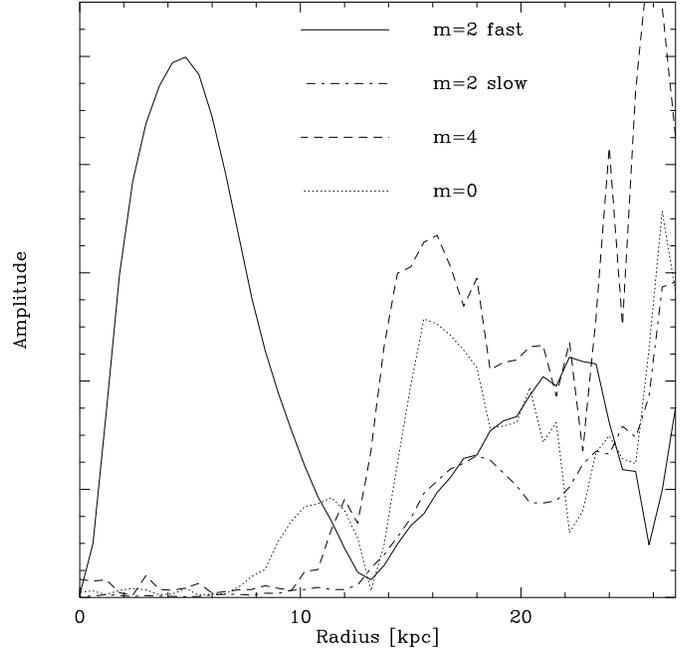,width=\columnwidth}
\caption{\label{fig:amplimode}
The  relative amplitudes of the modes, computed as explained in the
text. The ordinates scale is arbitrary, but is the
same for both $m=2$ on one hand, and for $m=4$ and $m=0$ on the other hand.
}
\end{figure}

The second question we have noted at the end of section~4.1
is whether, as expected from the theoretical works, the coupling
efficiency is very peaked at the bar corotation.  In order to answer
this question, we plot on Fig.~\ref{fig:amplimode} the amplitudes of
the modes involved in the coupling (the slow and fast $m=2$, the $m=0$
and the $=4$).  The amplitudes are computed by integrating, for each
value of $r$, the amplitude spectrum on a 1.6~km/s/kpc
bandwidth centered on the peak frequency of the mode.

\new{On this figure we clearly see from the solid line that, in
agreement with linear theory, the separation
between the bar and the Swing triggered spiral occurs at 13~kpc, \ie almost
exactly where the corotation appears to be located according to
Fig.~\ref{fig:run1_power_m2}.  }

For $r>13$~kpc the Swing-triggered spiral and the slower, non-linearly
excited one grow very similarly up to 18~kpc.  One can note that the
amplitudes of the $m=0$ and $m=4$ waves are clearly not proportional
to the product of the amplitudes of the fast and slow $m=2$ spirals:
in that case (where these waves could be simply understood as ordinary
beat waves, rather than partners in a non-linear mechanism), the $m=0$
and $m=4$ curves would peak around~18~kpc, and they would have a
parabolic shape between 13 and 18~kpc, since both $m=2$ curves are
linear on this range.  Instead of this behavior, both curves raise
sharply around 14~kpc and peak around 16~kpc.  This is a strong
indication that the coupling is very localized; indeed, just as the
slow $m=2$, the $m=0$ and $m=4$ are generated at this coupling radius,
and then propagate freely in the disk.  \new{Hence the coupling
partners can coexist on a wide range of radii whereas the non-linear
coupling mechanism which makes them interchange energy and angular
momentum takes place at a very well defined radius.}

Around 18-19~kpc, the $m=4$ is attenuated.  This is reasonable since
it reaches there its corotation, and thus the ``forbidden band'' where
it does not propagate (just as the fast $m=2$ at 14 kpc); on the other
hand the $m=0$ is also attenuated, something we could not expect from
its linear properties.  This can be understood by returning to the
energy spectrum of the $m=4$, Fig.~\ref{fig:run1_power_m4}: one
sees, at ~$30$~km/s/kpc, another quite strong $m=4$ feature with its
ILR at the corotation of the previous one; this is quite close to the
~$26$~km/s/kpc, expected for the beat wave of the previous $m=4$ and of
the $m=0$; we believe that non-linear coupling is at work also here,
allowing the $m=0$ and the fast $m=4$ to transfer their energy to a
slower one. This would be an illustration of the ``staircase'' of
modes often observed in these simulations.

\begin{figure}
\psfig{file=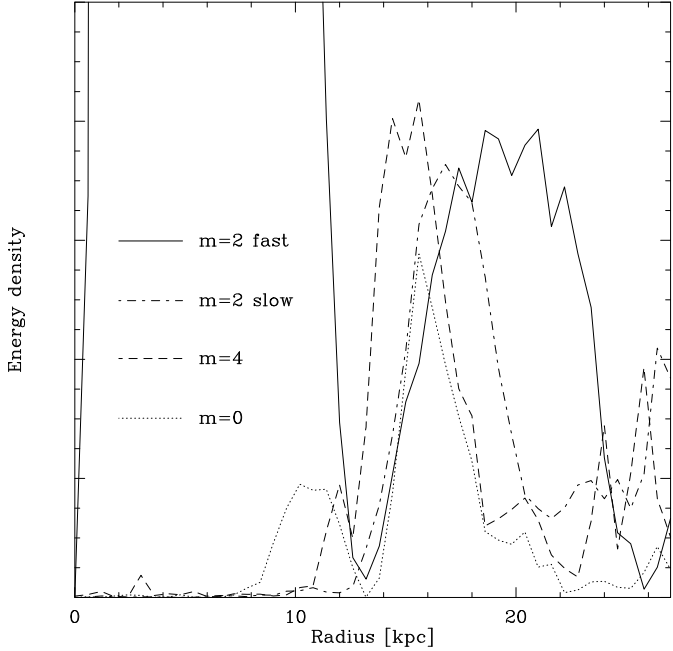,width=\columnwidth}
\caption{\label{fig:enermode}
This figure shows the mode energy density, computed in a similar
way as the amplitude at the previous figure (in particular over the
same bandwidth). Once again the vertical scale is arbitrary, but is the
same for both $m=2$ and for both $m=0$ and $m=4$.}
\end{figure}

Finally, one could wonder why the fast $m=2$ seems to extend (and even
to be peaked) beyond its OLR. Actually this behavior disappears if we
consider the energy density rather than the amplitude, as depicted on
Fig.~\ref{fig:enermode}.

On this figure we see that the fast $m=2$ spiral peaks around 20~kpc,
where its OLR is expected.  There is still a noticeable energy density
up to 2~kpc farther.  This is in agreement with the epicyclic radius
(which is the natural width of the resonance) of about
$\sigma_r/\kappa \sim 3$~kpc.  One can note once again on this plot
that the $m=4$ extends only to its corotation at $r \simeq 18$~kpc,
and that the $m=0$ also ends in this region, while the slow $m=2$
extends to $\sim 21$~kpc.

\subsection{Varying parameters in Run~1}
\new{In order to check the influence of the parameters of run~1 on the
spectra and on the disk profile temperature (which might be strongly
influenced by two-body relaxation in a 2D simulation), we have
repeated run~1 with twice as many particles, or with the same number
of particles but with a timestep of $0.25$~Myr, three times smaller
than the initial one.\\
\indent When we perform run~1 with a double number of particles, the
same features remain, \ie the bar, the Swing triggered spiral and the
slower spiral, at frequencies which have varied by no more
than~1~km/s/kpc.  The coupling partners also remain, with the same
intensities.  As expected, the outer part of the spectra is less noisy
since the particle noise has decreased.  Furthermore, the intermediate
faint $m=2$ standing mode of figure~\ref{fig:run1_power_m2} at
$22$~km/s/kpc almost disappears.  This is understandable since a good
explanation for this mode is the subharmonic excitation of the $m=4$
coupling partner, and subharmonic excitation requires initial noise to
start.  Since the noise is decreased in this new run, the
corresponding mode is accordingly fainter.  The same kind of argument
stands for the alternative mechanism one could consider to justify
this mode, which is the groove or ridge mechanism of Sellwood \&
Lin~1989.  Whatever the correct explanation for this mode may be, the
non-linear coupling mechanism which excites a slower spiral at its ILR
and at the bar corotation is far more relevant to account for the dynamics
of the external part of the galaxy.\\
\indent Let us mention that with twice as many particles, the
temperature profile obtained in the disk coincides with the one
observed with the initial number of particles (\ie~$80,000$).  This
shows that the heating in the simulations is due to the presence and
growth of the spiral and the bar, and not to two-body relaxation.\\
\indent When we perform run~1 with a smaller timestep ($0.25$~Myr
instead of $0.75$~Myr), we obtain the same results (\ie the same
features on all the spectra, with frequencies that have varied by no
more than~$0.5$~km/s/kpc), showing that the timestep chosen for
run~1 was sufficiently small.}

\subsection{Discussion about Run~1}

\subsubsection{Non-linear coupling versus grooves}

Run 1 confirms the coincidence of the corotation of the inner (fast) mode and
the ILR of the outer (slow) one, as noticed in the simulations of Sellwood
1985 by Tagger \etal 1987 and Sygnet \etal 1988.  This differs from
the mechanism of Sellwood \& Lin 1989, from which one would expect to
have the corotation of the outer mode at the OLR of the inner one:
Fig.~\ref{fig:run1_power_m2}
allows to clearly rule out this mechanism here, since the OLR of the
fast mode is at about 20~kpc, whereas the corotation of the slow
one is at about 25~kpc.  The run also confirms the
presence of coupled $m=0$ and $m=4$ waves at significant amplitudes.

Since the slower spiral is long-lived (as shown by the thinness of the
isocontours in Fig.~\ref{fig:run1_power_m2})
although it has an ILR where it should be
damped, it must be continuously fed.  We attribute this, following the
analysis of Tagger \etal 1987 and Sygnet \etal 1988, to non-linear
coupling between the two $m=2$ modes and the $m=0$ and $m=4$ beat
waves.

We also wish to mention the intermediate faint spiral mode at
$\omega=22$~km/s/kpc, observed on Fig.~\ref{fig:run1_power_m2},
whose corotation roughly coincides with the OLR of the fast spiral.
This mode could then correspond to a groove or ridge excited mode,
following the mechanism of Sellwood \& Lin 1989.  However the energy
flux it carries is far lower than the one transported by the fast and
slow spirals. The excitation of this mode, which appears
nearly halfway in frequency between the fast and slow spirals, could
have other explanations:

\begin{itemize}
\item{it could correspond to the subharmonic of the $m=4$ beat wave,
since it has half its wavenumber and nearly half its frequency. It would
be excited at its corotation where both waves are resonant.
}
\item{its OLR coincides with the corotation of the slow spiral, so
that it could be non-linearly fed by this slow wave.  In favor of this
explanation, one can notice that the low frequency, outer feature
observed on the $m=0$ power spectrum (slightly below $10$~km/s/kpc) would be an
obvious partner in this coupling.}
\end{itemize}

A combination of these three mechanisms is probably at work in the
generation of this intermediate $m=2$ feature, and much more detailed
simulations, with some additional theoretical work, would be needed to
understand their interplay. This might be taken as an illustration of
the fact that, at lower amplitudes than the dominant features we have
analyzed, one enters in the regime of multiple non-linear interactions
between numerous partners, which could lead in other conditions to a
turbulent cascade, as discussed in the introduction.

\subsubsection{Comparison of Swing and non-linear coupling}

An important difference between run 1 and the simulation of Sellwood
1985 (and many of our own runs) is that in the latter case the fast
mode did not extend beyond its corotation, whereas here the
Swing-triggered spiral (at the frequency of the bar) is nearly as
strong as the slower one up to 18~kpc, which corresponds to the
corotation of the $m=4$ coupled wave, and then even dominates it.  The
reason is that for this first run, in order to simplify the physics
involved, we have chosen parameters which optimize the initial
efficiency of the Swing, resulting in a very strong bar. Also, since
Sellwood's work clearly documents a case where the bar does not extend
beyond corotation, we have chosen this run here so as to show that
both behaviors are possible.

%

\begin{figure}
\psfig{file=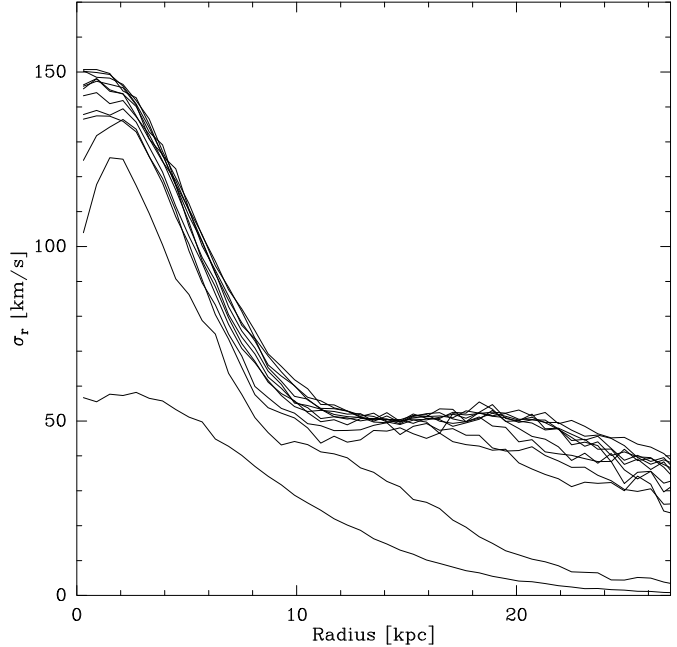,width=\columnwidth}
\caption{\label{fig:run1_sigmar}
This figure shows the radial velocity dispersion as a function of radius
for times 0, 1, 2\ldots 11~Gyr.  The lowest curve is for $t=0$,
the intermediate one is for $t=100$~Myr and the other ones almost coincide.
Since $Q \equiv 1.3$ at all
radii initially, one sees that, for $t > 100$~Myr
and $r \sim 15$~kpc, $Q$ is far above one, strongly limiting the
efficiency of non-linear coupling.}
\end{figure}

We also note, in reference to run 1, that the coupling efficiency,
which increases when the group velocities decrease as explained in
section~\ref{sec:localization}, is expected to vary as
$\sigma_r^{-3/2}$, where $\sigma_r$ is the radial velocity dispersion
(see Masset \& Tagger, 1996).  Since a realistic disk (\ie with a
dissipative component) would remain much colder than the disk of
run~1, we expect non-linear coupling to be far more efficient in a
realistic disk.  Fig.~\ref{fig:run1_sigmar} shows how the disk has
been heated by the bar.

We see that around the corotation of the bar,
where coupling occurs, the Toomre~Q parameter reaches values of 4 to 5
as soon as the bar forms.  For a colder disk, with $Q \sim 1.5$,
non-linear coupling could be up to 6~times more efficient.

On the other hand, the Swing mechanism also becomes less efficient as
$Q$ increases. Thus a realistic simulation including a gaseous
component and stars formation would be necessary to check
how the energetics of the Swing-triggered spiral and that of the
slower one compare in a realistic barred galaxy, as a function of
local or global parameters of the disk.

\section{Conclusion}

We have presented a typical numerical simulation of a barred spiral
galaxy, which shows a strong and unambiguous signature of non-linear
coupling between bar and spiral waves.  This non-linear coupling is
responsible for the excitation by the central bar of a slower, outer
$m=2$ spiral wave which can efficiently carry away the energy and
angular momentum extracted from the central regions by the bar. This
confirms, with more detailed diagnostics, the simulations of Sellwood
1985 and the theoretical explanation of Tagger \etal 1987, and
Sygnet \etal 1988.

This behavior is in fact routinely observed in numerical simulations,
as soon as one introduces realistically peaked rotation profiles at the
center, so that Lindblad resonances prevent any single spiral mode to
extend radially over most of the galactic disk. This leads us to
believe that non-linear coupling can be frequent also in real
galaxies, under different forms, which may be difficult to analyze
because of the complex density patterns (Sellwood \& Sparke 1988),
as discussed in the introduction:

\begin{itemize}
	\item {As mentioned by Sygnet \etal 1988, and in our introduction,
	SB(r) galaxies seem to be a very good candidate, because of the
	mismatch between the position angles of the bar and spiral found
	by Sandage (1961).  The disagreement of Buta (1987) on this
	observation might be attributed to the above-mentioned complex
	density patterns, so that this clearly deserves further
	investigations.}

	\item {As discussed by Sellwood \& Sparke 1988, this mechanism might
	help solve the long-standing difficulties found when one tries to
	locate the corotation in many barred spirals.}

	\item {Recent work based both on observations and more complex
	simulations (Friedli \& Martinet, 1993; \hbox{Friedli \etal, 1996}) has
	pointed to the frequent observation of ``bars within bars'' in the
	central regions of galaxies. They have shown that the inner bar is
	most frequently misaligned with the outer one, ruling out a purely
	dynamical origin (such as stars aligned on $x_2$ orbits,
	perpendicular to the main bar). Non-linear coupling, shown here to
	occur at larger radii, appears as a very good tentative explanation
	for this mechanism, which should play a major role in the fueling of
	the inner parts of the galaxy.}

	\item {M51 might provide a clue to the same mechanism in
	tidally-driven spirals.  Elmegreen \etal (1987), in detailed
	modeling of the spiral structure in three ``classical'' galaxies,
	found that the structure of M51 could be explained only by the
	presence of two distinct spirals, and noticed (before they knew about
	our work) that the corotation of the inner one would coincide with
	the ILR of the outer one -- the signature of our mechanism. One must
	naturally be cautious with this result, since the modeling of M51
	has proven to be a very challenging task. However, precisely because
	there remains much to be done in this respect, and because new
	observations become available, we believe that this might very well
	prove to be an important element in the complex physics of M51.}

	\item { There has been in recent years a renewed interest in $m=1$
	spiral structures.  The linear theory of the $m=1$ mode in
	galaxies, which differs markedly from $m>1$ ones, remains to be
	done, but it is generally believed that it is not or only weakly
	unstable.  Work done in the context of accretion disks (Adams
	\etal, 1989) would not really apply here, because self-gravity is
	strong and the boundary conditions used are different (see also
	Noh \etal, 1991).  Tagger \& Athanassoula (1990) discussed
	non-linear coupling of an $m=2$ with two $m=1$ modes, as a
	possible explanation for the structure of lopsided galaxies.  They
	appear as the strong version of a more frequent observation, that
	of off-centered nuclei (\ie nuclei affected by an $m=1$
	displacement) in galaxies - including our own.  Miller \& Smith
	(1992) for stellar disks, and Laughlin \& Korchagin (1996) for
	gaseous ones, have found persistent such motions in numerical
	simulations, a behavior we also obtain: although our cartesian
	grid gives insufficient resolution at the center to give it strong
	confidence without additional work, we do find $m=1$ spirals at
	the center, non-linearly coupled to unstable $m=2$ and $m=3$ ones.}

	\item {As mentioned in the introduction, we (Masset \& Tagger,
	1996b) have shown from analytical work that non-linear coupling
	with spiral waves is also a very tempting explanation for the
	generation of warps in disk galaxies. Work is in progress to give
	numerical evidence of this mechanism.}
\end{itemize}

\section{ Acknowledgments}
We wish to thank F. Combes and M. Morris for rich and helpful discussions
in the course of this work. We also thank A. Hetem for his close assistance
in software development, which has considerably increased its efficiency.
Finally we thank our referee D.~Friedli whose remarks and suggestions have
considerably improved the final version of this paper.

\end{document}